\documentclass[]{spie}

\usepackage{amsmath,amsfonts,amssymb}
\usepackage{graphicx}
\usepackage[colorlinks=true, allcolors=blue]{hyperref}
\usepackage{siunitx}

\title{The dependence of timing jitter of superconducting nanowire single-photon detectors on the multi-layer sample design and slew rate}
\author[a,*]{Rasmus Flaschmann}
\author[a]{Lucio Zugliani}
\author[a]{Christian Schmid}
\author[b]{Simone Spedicato}
\author[b]{Stefan Strohauer}
\author[a]{Fabian Wietschorke}
\author[b]{Fabian Flassig}
\author[b]{Jonathan J. Finley}
\author[a]{Kai M\"uller}

\affil[a]{Walter Schottky Institute and Department for Electrical and Computer Engineering, Technical University of Munich, 85748 Garching, Germany}
\affil[b]{Walter Schottky Institute and Physics Department, Technical University of Munich, 85748 Garching, Germany}

\authorinfo{*, Correspondence to rasmus.flaschmann@wsi.tum.de}

\begin{document}
\maketitle
\begin{abstract}
We investigated the timing jitter of superconducting nanowire single-photon detectors (SNSPDs) and found a strong dependence on the detector response.
By varying the multi-layer structure, we observed changes in pulse shape which are attributed to capacitive behaviour affecting the pulse heights, rise times and consequently timing jitter.
Moreover, we developed a technique to predict the timing jitter of a single device within certain limits by capturing only a single detector pulse, eliminating the need for detailed jitter measurement using a pulsed laser when a rough estimate of the timing jitter is sufficient.
\end{abstract}

\section{Introduction}

Recent years have seen major advances in photon based quantum technologies \cite{Kim08} such as deep space optical communication (DSOC) \cite{Cal19b, Iva20}, quantum key distribution (QKD) \cite{BB84, Shi14}, quantum computation \cite{Sin16}or state teleportation \cite{Pfa14}. 
For such applications, components such as single-photon emitters (e.g. quantum dots,\cite{Sen17} NV centres in diamond\cite{Kur00} or 2D materials\cite{Tra16}), spin-photon interfaces \cite{Ata18, Ber20} or single-photon detectors are needed.
In terms of detectors, superconducting nanowire single-photon detectors (SNSPDs) \cite{Gol01, Rei13, Red16, Fla22a} have prevailed over other options such as single-photon avalanche photodiodes (SPADs) \cite{Nix32} or transition edge sensors (TES) \cite{Ull15}.
Besides near-unity quantum efficiency\cite{Red20}, one of their outstanding performance features is the timing jitter describing the temporal resolution. \cite{Had09}
It was shown to be as low as \SI{2.6}{\pico\second} for a short (\SI{5}{\micro\meter} long) and thin (\SI{80}{\nano\meter} wide) superconducting nanowire \cite{Kor20} at $\lambda = $\SI{775}{\nano\meter}.
For a meandering nanowire covering an area of  10x\SI{10}{\micro\meter\squared}, values of \SI{28}{\pico\second} have been reported \cite{San19}.
Moreover, the timing jitter is a crucial parameter for various applications such as the characterization of single-photon emitters \cite{Had05}, single-photon QKD \cite{Kup20} or pulse-position modulation \cite{Shi99}, where information is encoded in so-called time bins.
The smaller the timing jitter the more time bins fit within a certain amount of time.
Therefore, it is of utmost interest to advance the understanding of the various influences on timing jitter, how to improve it, and how to speed up the characterization process.
To analyze the origin of timing jitter, it is useful to understand on which parameters it depends.
To this end, recent works have identified that the timing jitter depends on contributions from the amplifiers \cite{Wu17}, hotspot formation \cite{Kor17}, bias current and operation wavelength \cite{All19}, the pulse width of the laser, electrical and thermal \cite{San19} noise as well as the detector geometry \cite{Cal16}. \cite{San19}
Each of these components can be analyzed individually for instance by using a dual-readout scheme \cite{San19} and can be assigned to either intrinsic \cite{All19} (hotspot formation, detector geometry) or extrinsic (amplifiers, pulse width, noise) properties.
One parameter derived from the extrinsic properties is the signal-to-noise ratio (SNR), where You et al. were able to associate an improved jitter with an increased SNR \cite{You13}.
In addition, Wu et al. were able to demonstrate a direct correlation between the maximum edge slope (hereafter referred to as slew rate) and timing jitter \cite{Wu17}.
They compared different publications and found that lower jitter values were associated with a higher slew rate, which was confirmed by Korzh et al \cite{Kor20}.
In this work, we investigate this relation in more detail by analyzing the influence of different material layer combinations on the timing jitter and draw conclusions for optimized multi-layer structures for SNSPDs that enable high detection efficiency and low timing jitter.
Afterwards we show how to calibrate a measuring system to determine the jitter from the slew rate, i.e. the ratio of pulse height and rise time.

\section{Experimental}

\begin{figure}[h]
\centering
  \includegraphics[height=5cm]{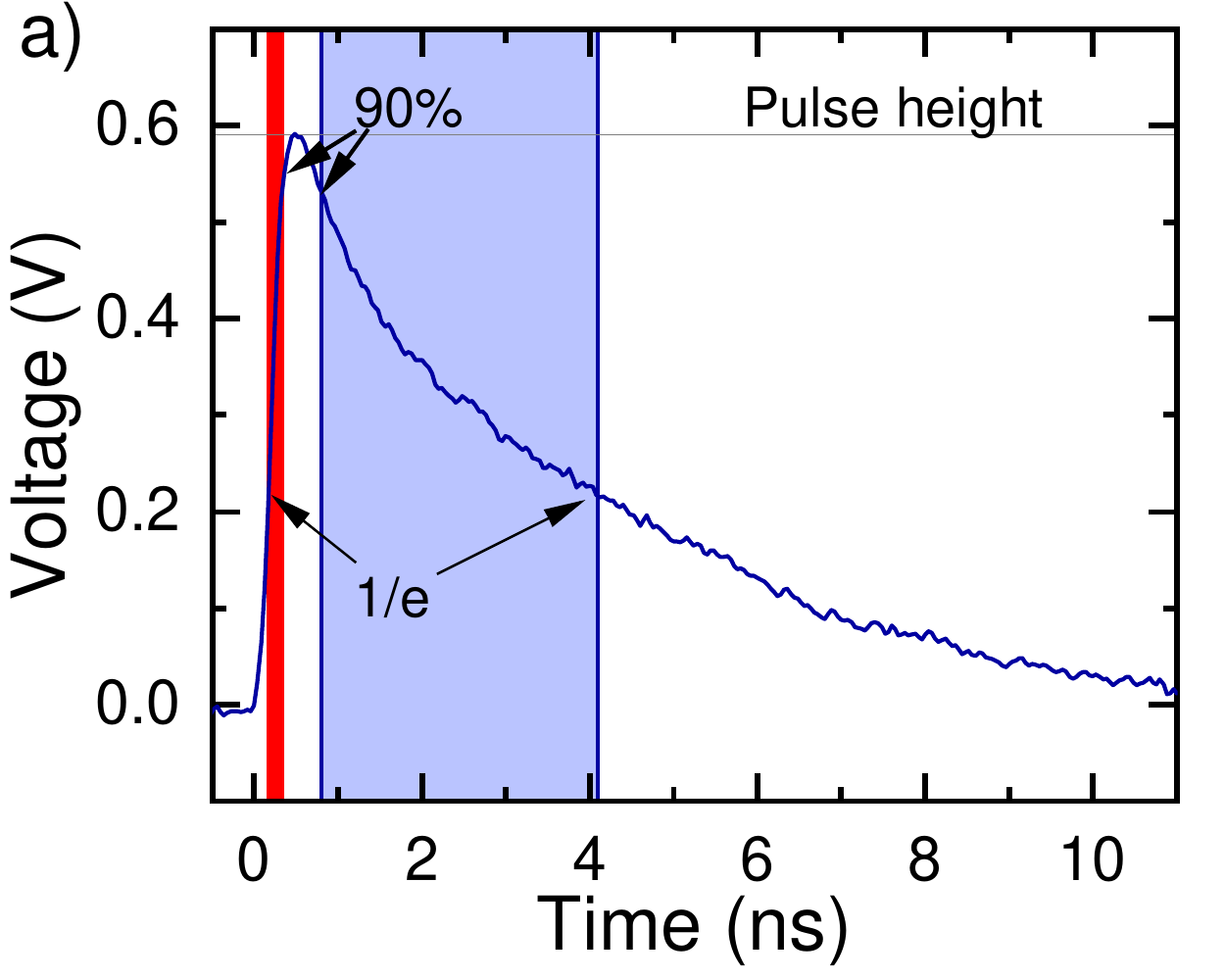}
  \includegraphics[height=5cm]{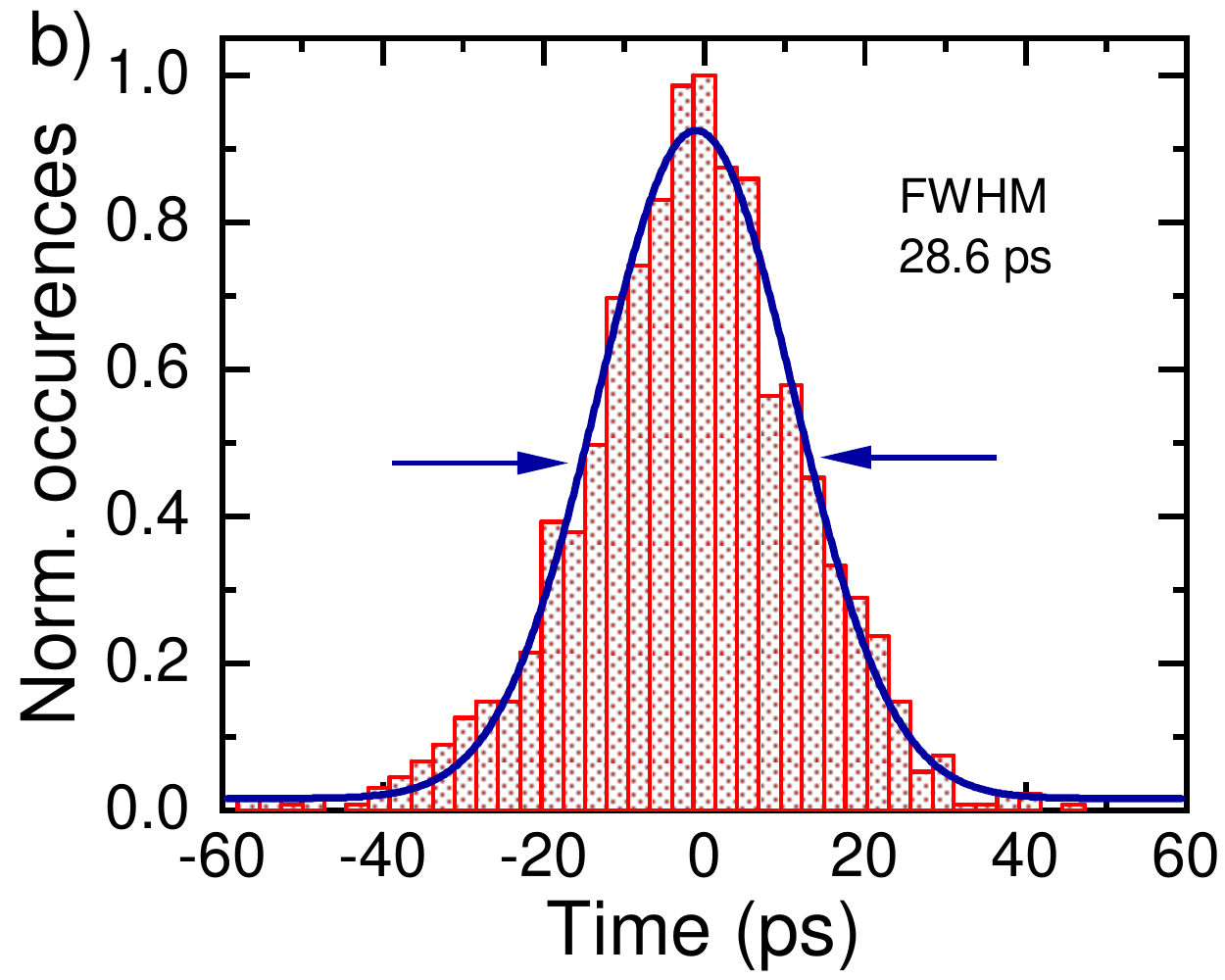}
  \caption{Detector response upon photon absorption. (a) Typical voltage pulse showing a fast rising edge and exponential decay. The rise time is defined as the time between 1/e and \SI{90}{\percent} of the maximum pulse height. (b) Timing jitter determined using a pulsed laser to perform a start-stop measurement between a photodiode (start signal) and the detector unit (stop signal). The resulting normalized occurrences are plotted as a function of time. The timing jitter is determined as the FWHM of the obtained Gaussian distribution.}
  \label{fig:1}
\end{figure}

All measurements were carried out with a Janis cryogenic probestation at \SI{4.5}{\kelvin}.
The detectors are made of niobium titanium nitride (NbTiN), with a typical thickness of \SI{8}{\nano\meter}, nanowire width of \SI{100}{\nano\meter} and an areal fill factor of \SI{33}{\percent} covering an area of 10x\SI{10}{\micro\meter\squared}.
For the efficiency measurements a continuous wave (CW) laser (\SI{780}{\nano\meter}) was used to illuminate the detectors.
In Fig. \ref{fig:1}(a) a typical detector response is shown.
It consists of a fast rising edge and an exponential decay following $\tau_\text{fall} = L_\text{k} / Z_\text{load}$ \cite{Ker06}, where $Z_\text{load}$ is the load impedance (typically \SI{50}{\ohm}) and $L_\text{k} = \frac{\hbar R_\text{device}}{\pi 1.76 k_\text{B} T_\text{c}}$ \cite{Yan17} the kinetic inductance, which was \SI{194}{\nano\henry} for the detectors throughout this study.
Here $R_\text{device}$ describes the device resistance and $T_\text{c}$ corresponds to the switching temperature of the superconductor.
For the timing measurements a pulsed laser (pulse width $<\SI{3}{\pico\second}$, at \SI{850}{\nano\meter}) was used.
The laser signal is divided between a ultrafast photodiode (UPD-15-IR2-FC, start signal) with a total rise time of less than \SI{15}{\pico\second} and the detector unit consisting of the detector and two room temperature amplifiers (combined \SI{53.5}{dB}, stop signal).
Both channels are connected to a sampling scope (MSO64) to perform a delay measurement.
In Fig. \ref{fig:1}(b), the normalized occurences of the delay measurement are plotted as a function of time.
The corresponding full width at half maximum (FWHM) of the obtained Gaussian distribution represents the devices timing jitter (here: \SI{28.6}{\pico\second}).

\begin{table}[h]
\centering
\small
  \caption{\ Timing jitter of the measurement setup, characterized by mimicking the detector signal with an arbitrary waveform generator before it is attenuated and amplified again. For the resulting pulse height, the timing jitter is measured as depicted in Fig. \ref{fig:1}(b) and drops from more than \SI{60}{\pico\second} at \SI{0.1}{\volt} over \SI{16.4}{\pico\second} at \SI{0.4}{\volt} to just under \SI{10}{\pico\second} at \SI{0.7}{\volt}.}
  \label{tbl:1}
  \begin{tabular*}{0.6\textwidth}{@{\extracolsep{\fill}}ll|ll}
    \hline
    Pulse height (V) & Jitter (ps) & Pulse height (V) & Jitter (ps) \\
    \hline
    0.1  & 61.1 & 0.4  & 16.4 \\
    0.15 & 41.9 & 0.45 & 14.4 \\
    0.2  & 31.6 & 0.5  & 12.6 \\
    0.25 & 26.4 & 0.55 & 11.6 \\
    0.3  & 22.0 & 0.6  & 10.8 \\
    0.35 & 19.0 & 0.7  & 9.6  \\
    \hline
  \end{tabular*}
\end{table}

To characterise the timing jitter of the setup, an arbitrary waveform generator (AWG) was used that mimics the detector signal before it is attenuated and re-amplified.
Apart from the signal source, the measurement setup matches that of the detector measurement.
Depending on the amplified pulse height we observe a drop from more than \SI{60}{\pico\second} at \SI{0.1}{\volt} over \SI{16.4}{\pico\second} at \SI{0.4}{\volt} to just under \SI{10}{\pico\second} at \SI{0.7}{\volt} (cf. Tab. \ref{tbl:1}).
Hence, the timing jitter of the setup (cf. Tab. \ref{tbl:1}), mainly originating from the room temperature amplifiers, is significantly smaller than the measured device jitter values (cp. Fig. \ref{fig:1}) ensuring that it did not influence the presented results.
This can be analyzed using $t_\text{tot} =  \sqrt{\Sigma t_\text{i}^2}$, where $t_\text{tot}$ is the measured timing jitter and $t_\text{i}$ are the different contributions.
The uncertainty of the data presented in Tab. \ref{tbl:1} is mainly limited by the internal device jitter of the sampling scope (\SI{1.5}{\pico\second}) and the fast photodiode approximately to be less than \SI{5}{\pico\second}.

\section{Results}
To investigate the dependence of the pulse shape on the timing jitter, samples with different multi-layer structures were fabricated.
While most detectors were produced on a full-chip sized (FCS) gold mirror (\SI{10}{\nano\meter} Ti / \SI{50}{\nano\meter} Au) with varying silicon dioxide (SiO$_2$) layer thicknesses to form a single-sided cavity, detectors were furthermore fabricated on small (\SI{100}{\micro\meter} diameter) gold mirrors (hereafter: Au$^{*}$) with \SI{105}{\nano\meter} SiO$_2$ and on top of \SI{130}{\nano\meter} SiO$_2$ on a silicon wafer (hereafter: Wafer+130).

\begin{figure}[h]
\centering
  \includegraphics[height=5cm]{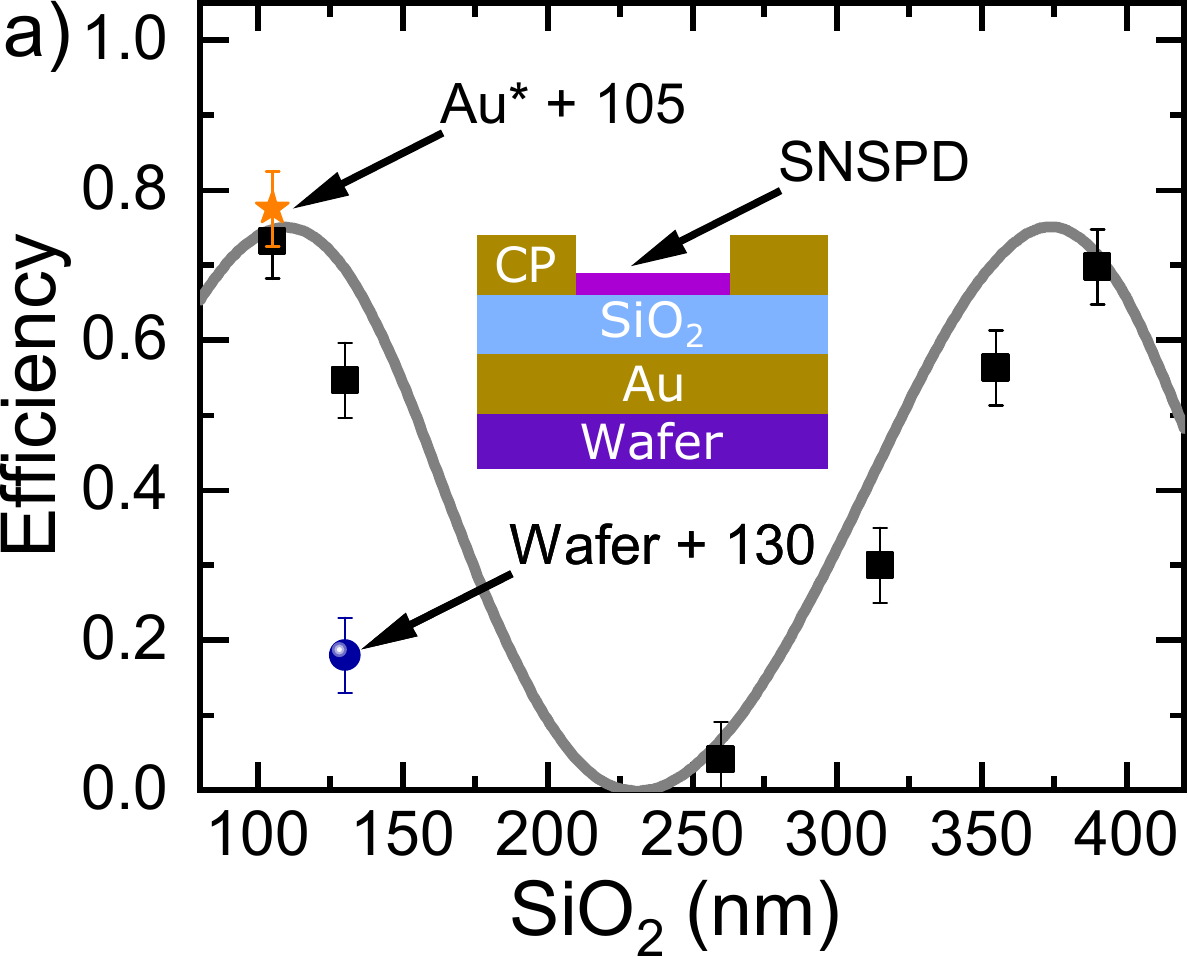}
  \includegraphics[height=5cm]{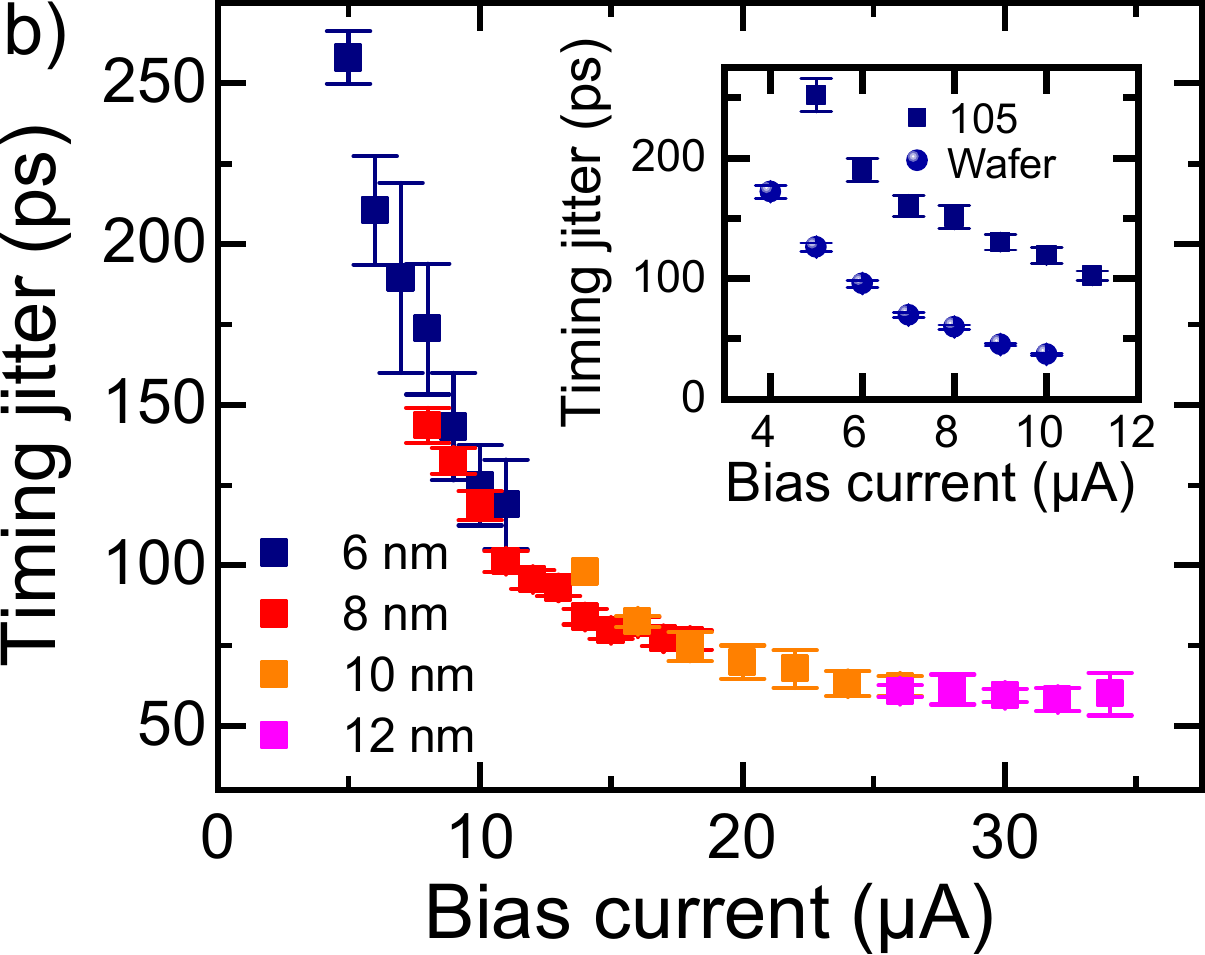}
  \caption{Influence of multi-layer structure. (a) Measured and simulated efficiency of \SI{8}{\nano\meter} NbTiN at \SI{780}{\nano\meter} on gold mirror with varied SiO$_2$ thickness. Additional data points show the efficiency on a \SI{100}{\micro\meter} diameter gold mirror (hereafter: Au$^{*}$) with \SI{105}{\nano\meter} SiO$_2$ and on top of \SI{130}{\nano\meter} SiO$_2$ on a silicon wafer (hereafter: Wafer+130). The measured efficiencies agree very well with the simulated values. The inset depicts the multi-layer structure used including the (\SI{10}{\nano\meter} Ti / \SI{50}{\nano\meter} Au) contact pads. (b) Timing jitter depending on the applied current using different NbTiN thicknesses on gold mirrors with a  \SI{105}{\nano\meter} SiO$_2$ layer on top and measured at a wavelength of \SI{850}{\nano\meter}. The jitter decreases with an increased current converging against \SI{65}{\pico\second}. The inset shows a comparison in the range from \SIrange{4}{12}{\micro A} of the same data with detectors directly fabricated on a Si/SiO$_2$ wafer. We observe that the jitter for the detectors on a Si/SiO$_2$ wafer are more than \SI{50}{\pico\second} lower compared to those on a full-chip sized (FCS) mirror with \SI{105}{\nano\meter} SiO$_2$ on top.}
  \label{fig:2} 
\end{figure}

Fig. \ref{fig:2}(a) shows the efficiency of the aforementioned samples as a function of the silicon dioxide thickness measured at a wavelength of \SI{780}{\nano\meter}.
In agreement with finite difference time domain (FDTD) simulations, the data reveal a maximum efficiency around \SI{75}{\percent} for \SI{105}{\nano\meter} SiO$_2$ (1$^\text{st}$ maximum) and \SI{390}{\nano\meter} SiO$_2$ (2$^\text{nd}$ maximum) on a full-chip sized gold mirror as a bottom cavity.
While similar results can be achieved for small gold mirrors with \SI{105}{\nano\meter} SiO$_2$, the efficiencies for the data points in between 
are significantly lower in agreement with FDTD simulations.
The inset depicts a sketch of the multi-layer structures used including the (\SI{10}{\nano\meter} Ti / \SI{50}{\nano\meter} Au) contact pads.
In Fig. \ref{fig:2}(b) the timing jitter is shown as a function of the applied bias current for different NbTiN thicknesses on a FCS gold mirror with \SI{105}{\nano\meter} SiO$_2$.
The jitter throughout this work was measured at a wavelength of \SI{850}{\nano\meter} and decreases with an increased current converging against \SI{65}{\pico\second}.
The inset shows the same data in comparison to detectors fabricated on a Si/SiO$_2$ wafer in the range up to \SI{12}{\micro A}.
Here, a significantly improved jitter can be observed for the detectors without Au mirror, reaching values of \SI{37}{\pico\second}.
Therefore, the comparison indicates a strong influence of the gold mirror underneath.

\subsection*{Influence of the multi-layer structures}

\begin{figure}[h]
\centering
  \includegraphics[height=5cm]{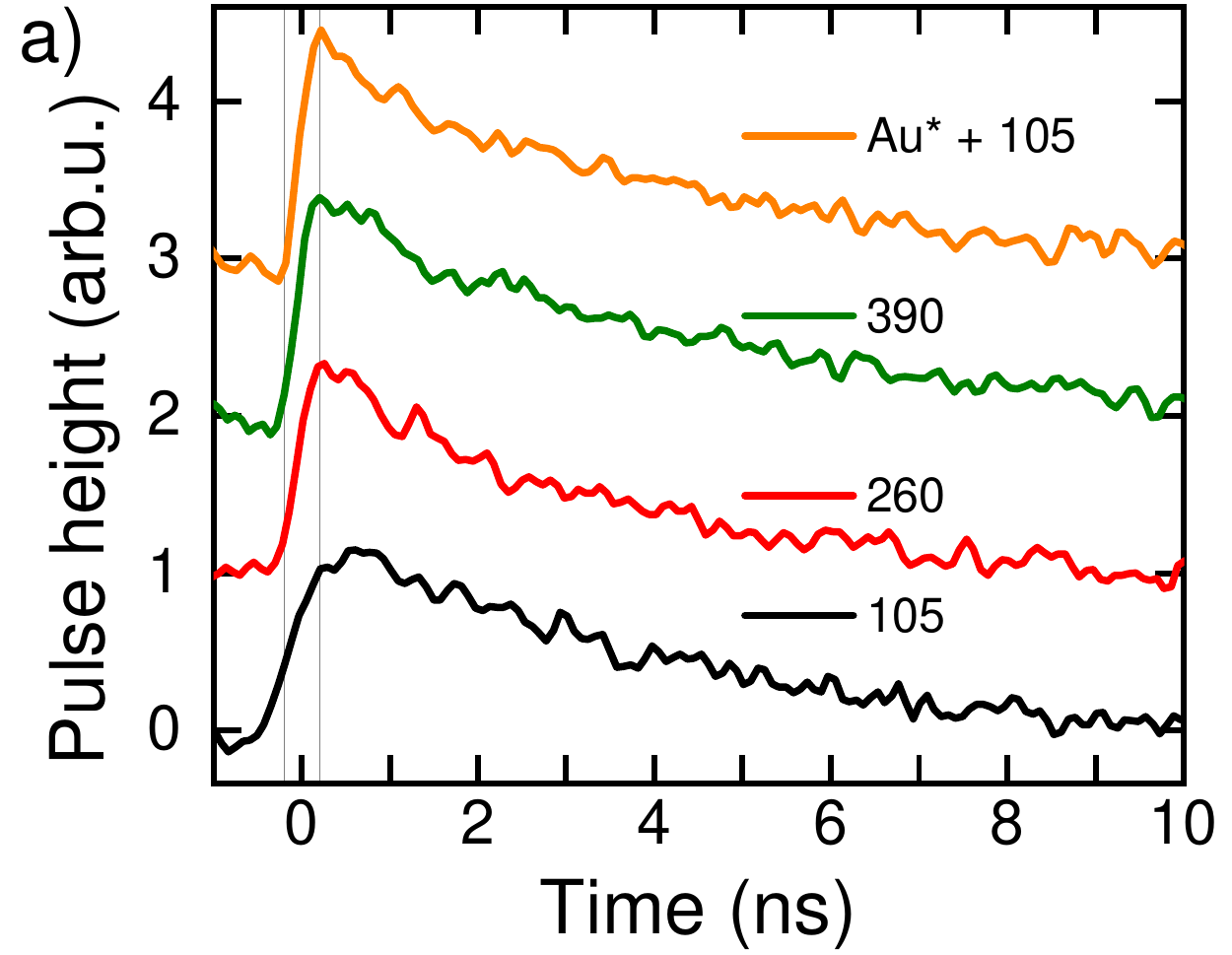}
  \includegraphics[height=5cm]{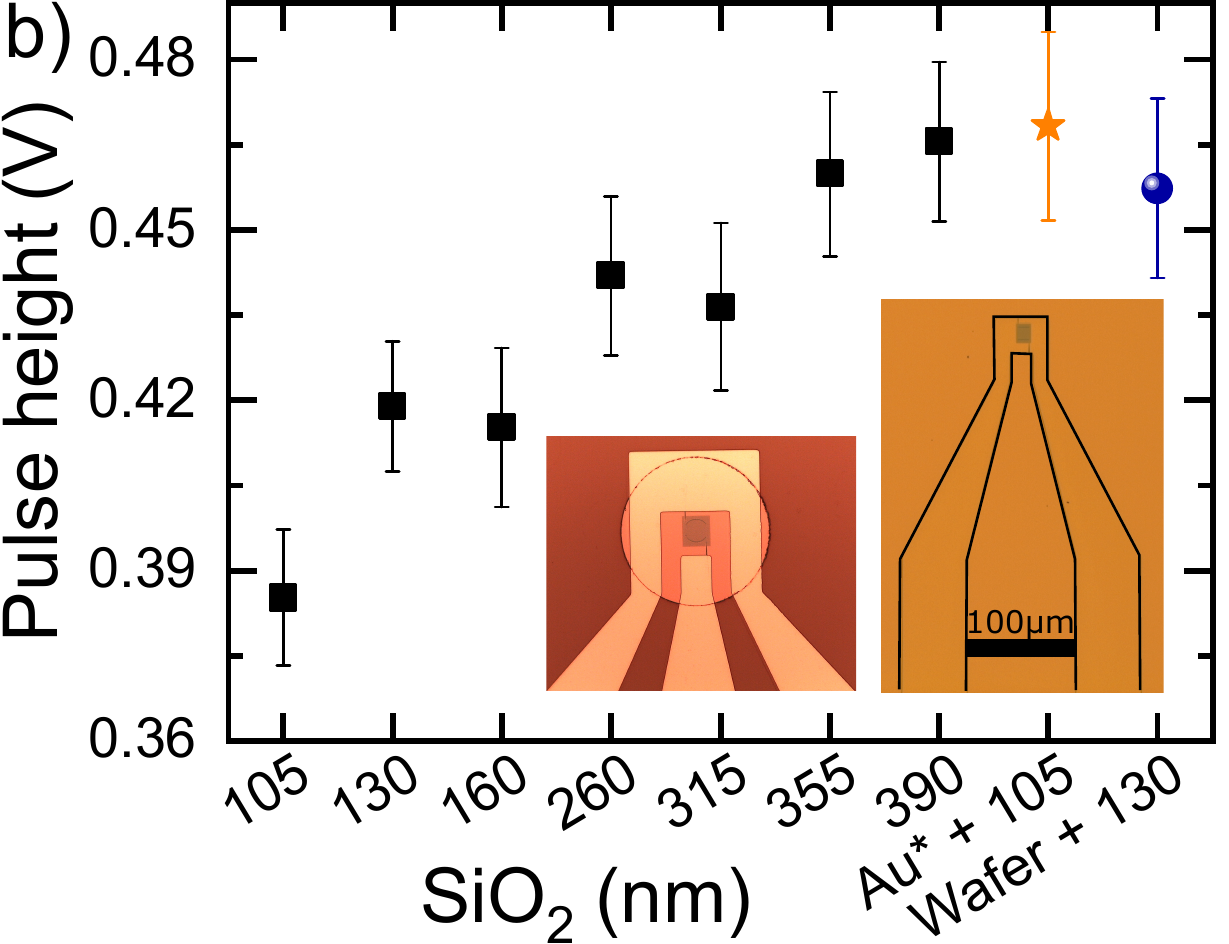}
  \caption{Comparison of the detector response for different multi-layer structures at an applied bias current of \SI{16}{\micro A}. (a) Normalized voltage pulses shown for \SI{105}{\nano\meter}, \SI{260}{\nano\meter}, \SI{390}{\nano\meter} SiO$_2$ on a FCS gold mirror and \SI{105}{\nano\meter} SiO$_2$ on a small gold mirror. The reference lines mark the start and end position of the rising edge for the \SI{100}{\micro\meter} mirror. The rise time decreases and the pulse height increases with an increasing SiO$_2$ thickness. (b) Mean pulse height for detectors fabricated on different material combinations. The pulse height increases with an increasing SiO$_2$ thickness and for the detectors fabricated on wafer and small gold mirror. The inset depicts microscope images of the detector on both gold mirror types.}
  \label{fig:3}
\end{figure}

\begin{figure}[h]
\centering
  \includegraphics[height=5cm]{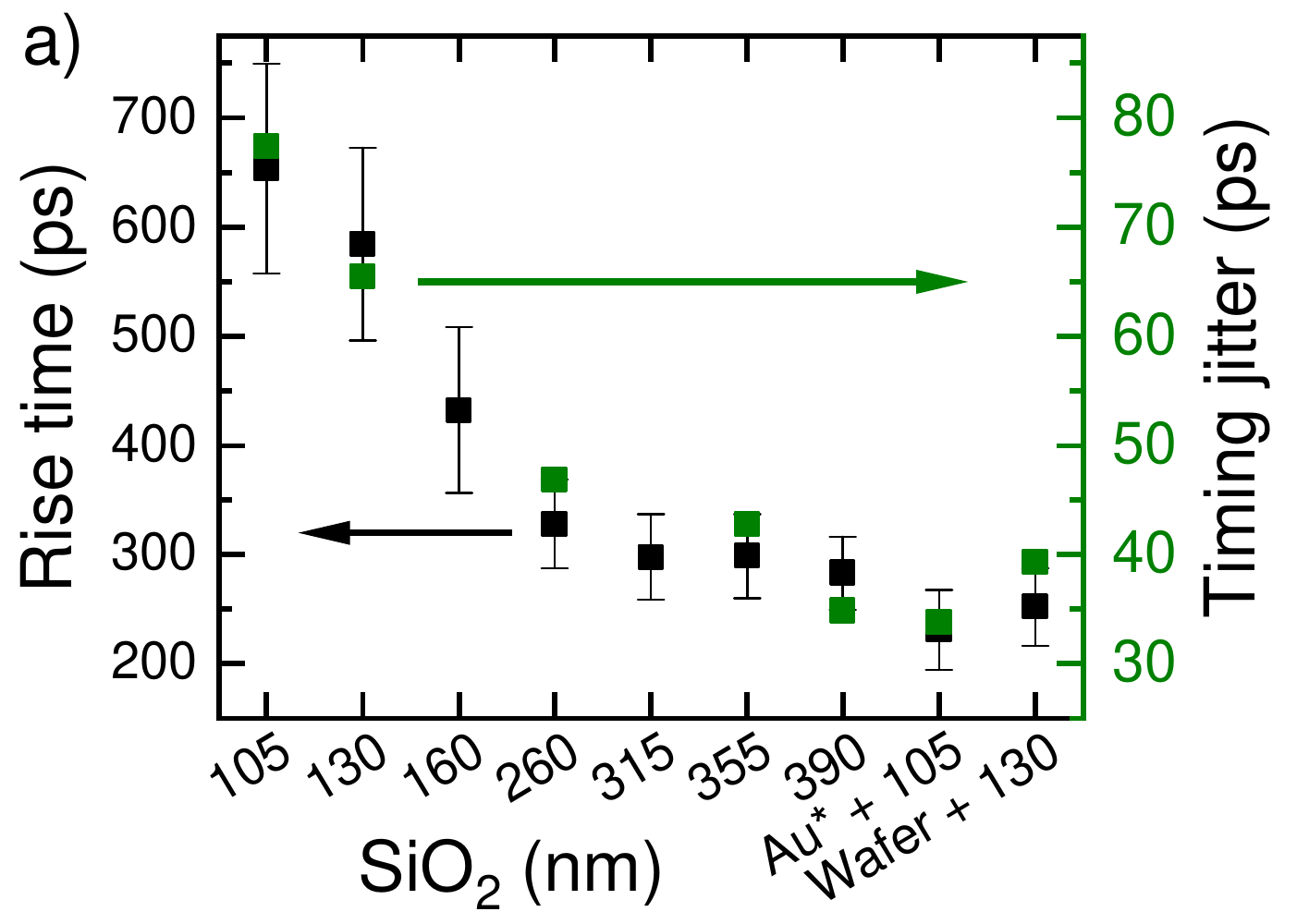}
  \includegraphics[height=5cm]{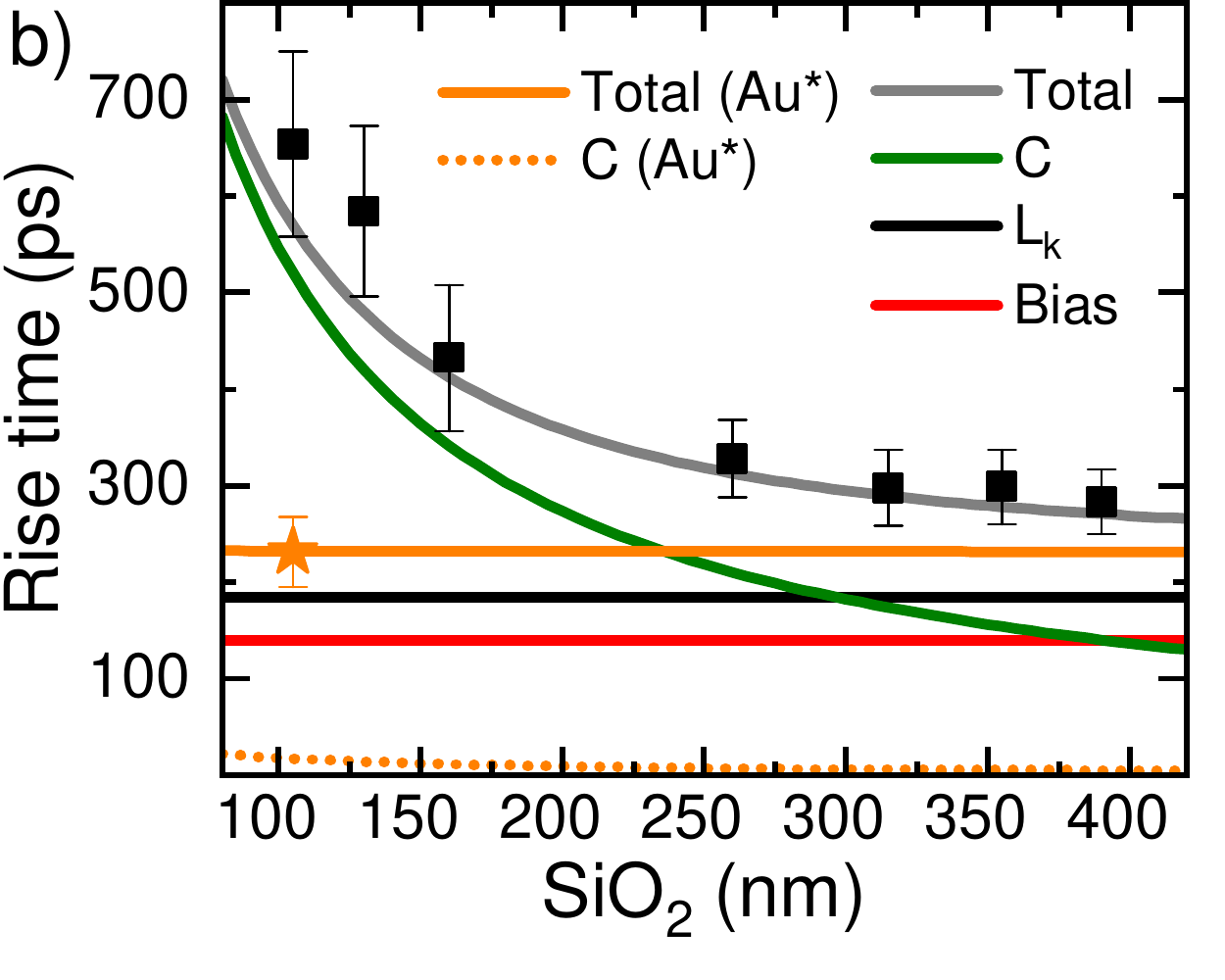}
  \caption{Influence of distance and size of the buried gold mirror relative to the contact pad at a bias current of \SI{16}{\micro A}. (a) Rise time and timing jitter for the different multi-layer structures used. We observe a parallel decrease of both the rise time and jitter with an increased SiO$_2$ thickness or small gold mirrors. While the rise time decreases from \SIrange{659}{231}{\pico\second}, the jitter improves from \SIrange{77}{33}{\pico\second}. (b) Calculated rise time consisting of a capacitance (C), bias current (Bias) and kinetic inductance (L$_\text{k}$) component. While the kinetic inductance and bias component are the same for all material combinations, an additional capacitance contribution is formed between the contact pads and the FCS gold mirrors in agreement with the measured rise times.}
  \label{fig:4}
\end{figure}

To analyze the influence of the multi-layer structures, detector pulses  for \SI{105}{\nano\meter}, \SI{260}{\nano\meter}, \SI{390}{\nano\meter} SiO$_2$ on a FCS gold mirror and \SI{105}{\nano\meter} SiO$_2$ on the small gold mirror are shown in Fig. \ref{fig:3}(a).
The pulses exhibit an increased steepness (from bottom to top) resulting not only in a shorter rise time but also in higher pulses.
Note that the pulse for the \SI{105}{\nano\meter} SiO$_2$ on a small gold mirror shows a significantly steeper rising edge compared to the same SiO$_2$ thickness on a FCS film.
Moreover, for a FCS film a general trend of faster rise times for thicker SiO$_2$ layers can be observed.
In Fig. \ref{fig:3}(b) we present the corresponding pulse height as a function of the multi-layer structure used at a fixed bias current of \SI{16}{\micro A}.
From \SI{0.39}{\volt}, the pulse height increases with an increasing SiO$_2$ thickness until it reaches the same pulse height of about \SI{0.47}{\volt} as both the detector on the small gold mirror (inset Fig. \ref{fig:3}(b)) and on Si/SiO$_2$ wafers.
To conclude, for thicker SiO$_2$ thicknesses (or small gold mirrors) the rise time becomes shorter and the pulse height larger.
To quantify these findings, we plot the rise time and timing jitter for the different samples at a fixed bias of \SI{16}{\micro A} in Fig. \ref{fig:4}(a).
Here, the timing jitter follows the same trend as the rise time.
While for \SI{105}{\nano\meter} SiO$_2$ on an FCS film a timing jitter of only \SI{80}{\pico\second} was measured, we achieved around \SI{30}{\pico\second} for the same SiO$_2$ thickness on a small mirror.
Again, the detectors on \SI{390}{\nano\meter} SiO$_2$/Au, a small gold mirror and the detectors directly fabricated on the wafer behave similar confirming the results in Fig. \ref{fig:3}.
Fig. \ref{fig:4}(b) depicts the rise time as a function of the SiO$_2$ thickness for different multi-layer structures.
The bias (red) and kinetic inductance component (black) are constant at a fixed bias level (here: \SI{16}{\micro A}) and the same for both types of mirrors.
The capacitance component on the other hand increases with an increasing SiO$_2$ thickness, but also strongly depends on the mirror design (dashed orange and solid green lines for Au$^{*}$ and full-chip size mirrors, respectively).
The resulting total rise times for Au$^{*}$ mirror (solid orange line) and full-chip size mirror (solid gray line) can be calculated using $\tau_\text{tot} = \sqrt{\tau_\text{Lk}^2 + \tau_\text{bias}^2 + \tau_\text{C}^2}$ \cite{Val48}.
The fraction of the kinetic inductance ($\tau_\text{Lk} = \alpha \cdot L_\text{k} / R_\text{n}$ \cite{Smi16}) depends on the scaling factor $\alpha = 0.9-1/\text{e}$, which relates to the limited rising edge from 1/e to \SI{90}{\percent}, the kinetic inductance, and the resistance of the normal-conducting $R_\text{n}$ region upon photon absorption.
The normal conducting resistance can be calculated as $R_\text{n} = \frac{V_\text{HS}}{V_\text{SNSPD}} \cdot R_\text{device}$.
With the hotspot volume $V_\text{HS}$ \cite{Sem09}, the SNSPD volume $V_\text{SNSPD}$ depending on the device geometry and the measured resistance $R_\text{device}$ of the presented detectors we determine it to be $R_\text{n} = \SI{0.56}{\kilo\ohm}$.
The second rise time component is related to the bias current.
It decays with $\tau_\text{bias} \propto \sqrt{1 / I_\text{Bias}}$\cite{Nic19} (cp. Fig. \ref{fig:5}(a)) but stays constant at a fixed bias.
To derive it, we considered detectors on a small mirror with a known kinetic inductance component.
The third parameter is a capacitive component forming between the FCS (small) gold mirror and the gold contact pads.
For each type of mirror we assume a plate capacitor $ \frac{1}{C_\text{tot}} = \frac{1}{C_\text{Sig}} + \frac{1}{C_\text{Gnd}} = \frac{d_\text{SiO2}}{\epsilon_\text{0} \epsilon_\text{R}} (\frac{1}{A_\text{Sig}} + \frac{1}{A_\text{Gnd}})$ considering both contributing areas signal ($A_\text{Sig}$) and ground ($A_\text{Gnd}$) separately, with $d_\text{SiO2}$ as the thickness of the silicon dioxide layer, $\epsilon_\text{0}$ as the vacuum permittivity and $\epsilon_\text{R}$ as the relative permittivity of silicon dioxide.
Subsequently, we calculate the cutoff frequency $f_\text{cutoff} = \frac{1}{2\pi C_\text{tot} Z_\text{load}}$ and get the capacitive component of the rise time $\tau_\text{C} = \frac{1}{2f} \cdot \alpha$.
The difference in $\tau_\text{C}$ (orange dashed vs. green line) shown in Fig. \ref{fig:4}(b) is due to the different areas contributing to the capacitance.
Thus, by choosing the small gold mirror, the rise time and therefore timing jitter can be improved significantly.

\subsection*{Current dependence of rise time, jitter and pulse height}

\begin{figure}[h]
\centering
  \includegraphics[height=5cm]{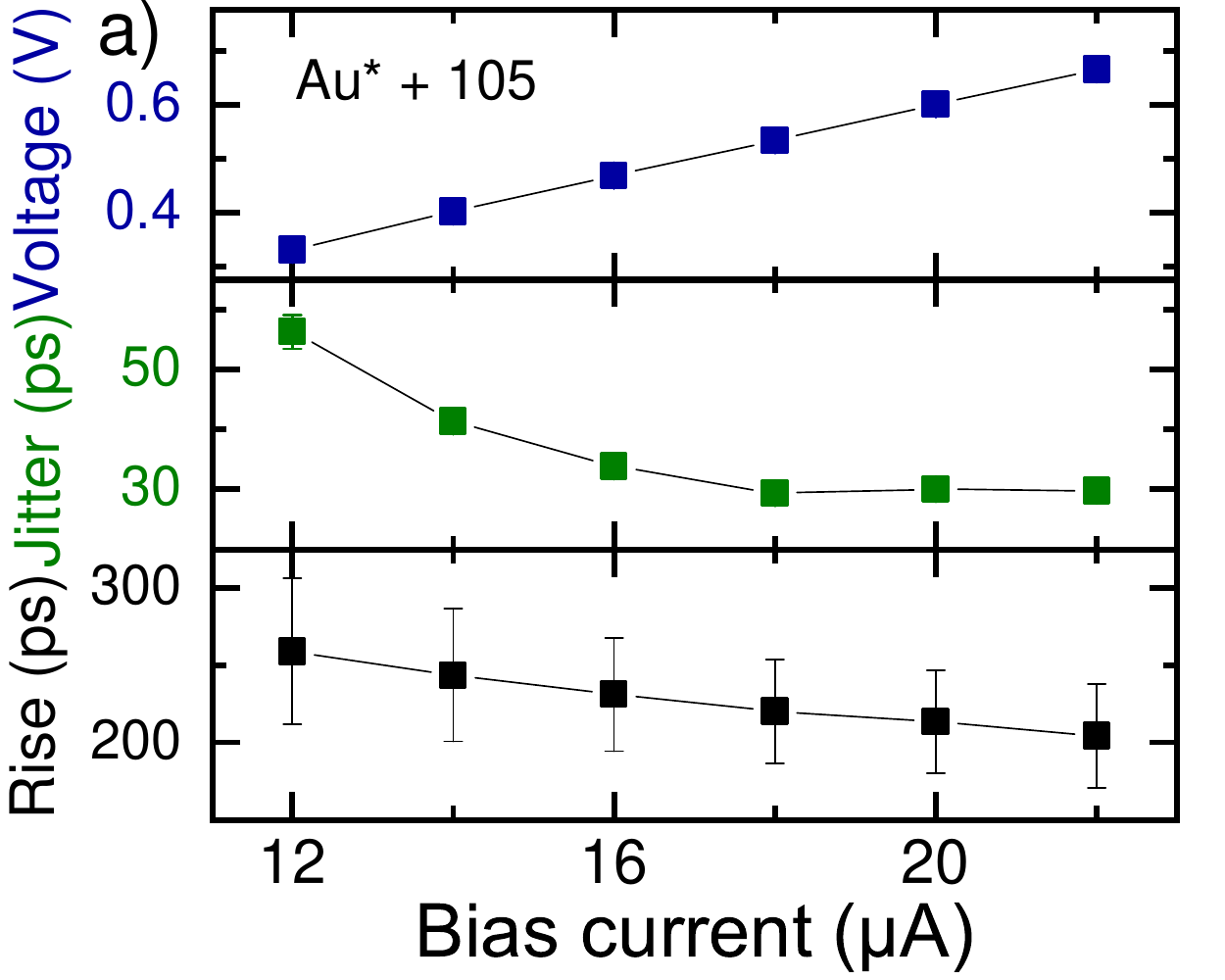}
  \includegraphics[height=5cm]{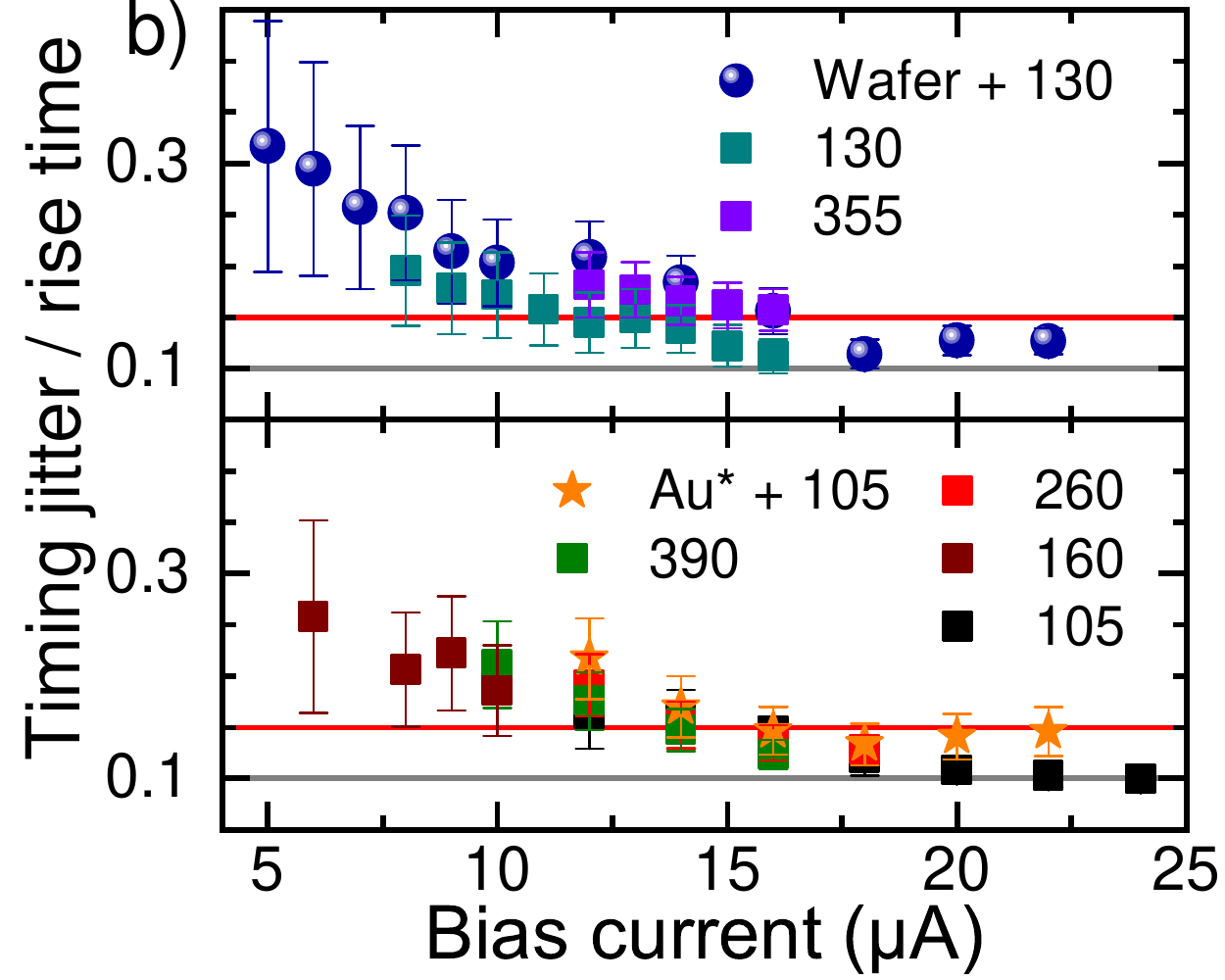}
  \caption{Current dependence of different pulse parameters. (a) Current dependence of the pulse height (top), the timing jitter (middle) and the rise time (bottom) of the detector on a gold mirror with a diameter of \SI{100}{\micro\meter}. While the pulse height increases linearly, both the rise time and jitter decrease with an increasing bias current. (b) Ratio of rise time and timing jitter as a function of the applied current. Due to the large number of material combinations, the data set was divided between the bottom and top panel. The overall trend shows a decrease of the ratio from 0.3 at \SI{5}{\micro A} down to 0.1 above \SI{20}{\micro A} regardless of the material layer combination and only depending on the applied current.}
  \label{fig:5}
\end{figure}

After the impact of the multi-layer structure, we now consider the current dependence on the timing characteristics.
To this end Fig. \ref{fig:5}(a) shows a typical detectors dependence of the rise time, timing jitter and pulse height on the applied bias current.
The data reveals a linear increase for the pulse height, an exponential decrease in the timing jitter (compare Fig. \ref{fig:2}(b)) and a slow decrease of the rise time.
Considering the latter two, we present the ratio of timing jitter and rise time as a function of the applied bias current in Fig. \ref{fig:5}(b).
The data was divided between the top and bottom panel due to the large number of material combinations.
Interestingly, all data points of the different multi-layer structures follow the same trend.
This is very surprising as the rise time in Fig. \ref{fig:4}(a) differs strongly for the different samples.
However, it can be concluded that the ratio of timing jitter and rise time is stable for a fixed bias current (within a current dependent limit) regardless of the material combination used.
This is an important finding for the characterization of detectors, as it allows to approximate the jitter for a given bias current and rise time.

\subsection*{Ratio of timing jitter and slew rate}

\begin{figure}[h]
\centering
  \includegraphics[height=5cm]{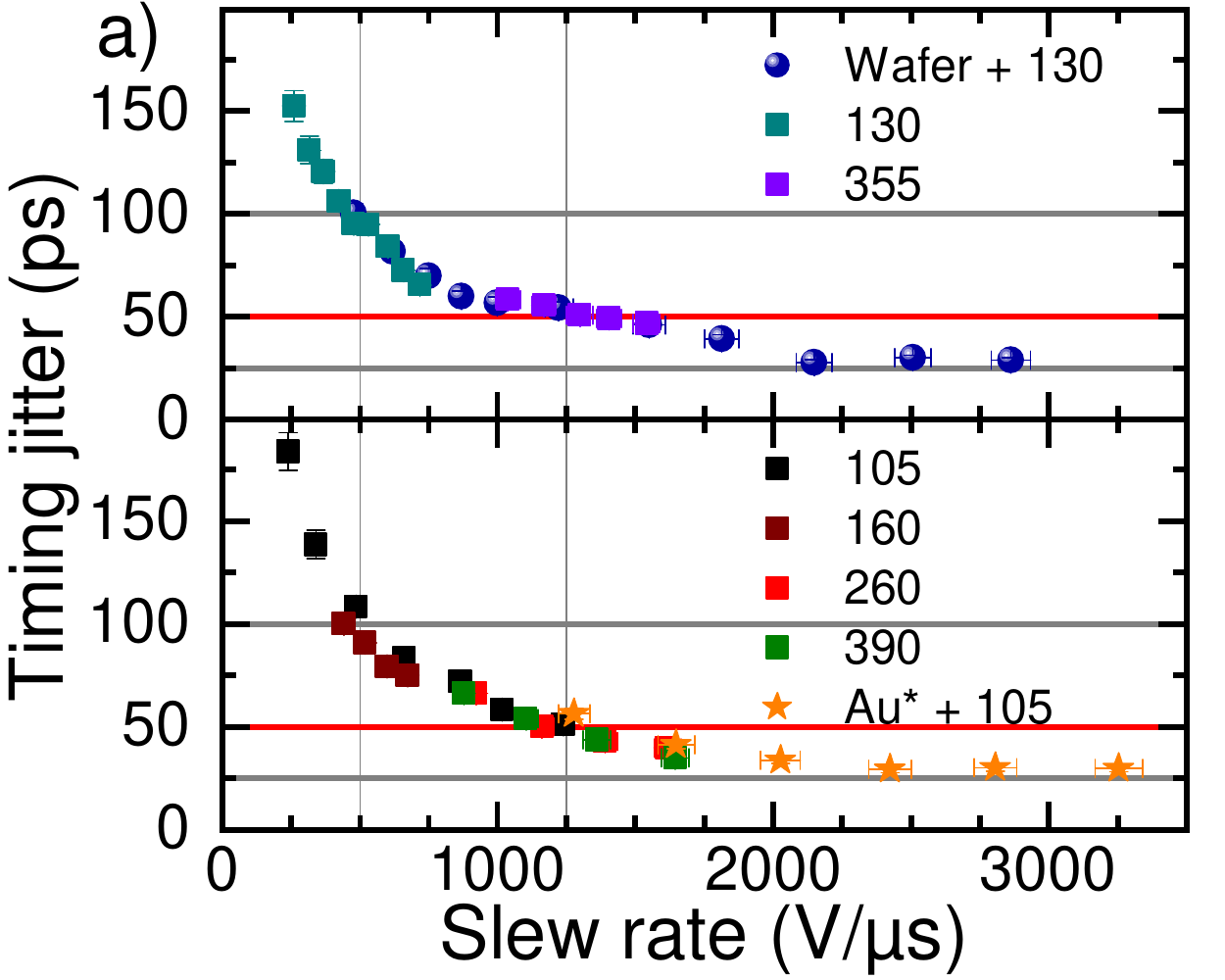}
  \includegraphics[height=5cm]{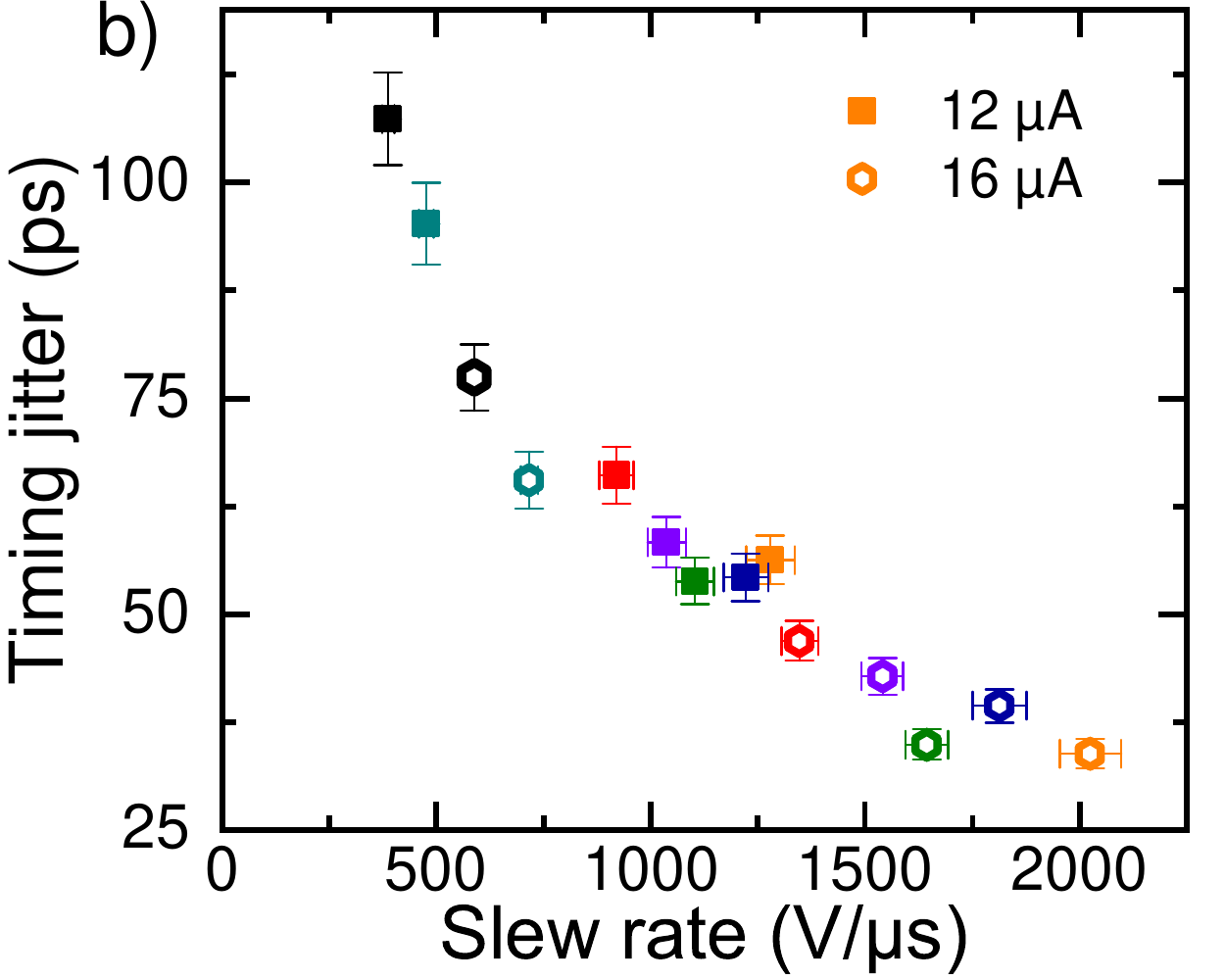}
  \caption{Timing jitter as a function of the slew rate. (a) Due to the large number of material combinations, the data set was divided between the bottom and top panel. Both curves show the same trend of an improved jitter with an increasing slew rate, approaching a jitter of \SI{25}{\pico\second}. Reference lines at \SI{500}{\volt / \micro\second} and \SI{1250}{\volt / \micro\second} show a timing jitter of \SI{100}{\pico\second} and \SI{50}{\pico\second}, respectively, regardless of the material combination used. (b) Dependence of the timing jitter on the slew rate for different bias currents. The slew rate increases with an increased bias current at constant SiO$_2$ thickness. Furthermore, the slew rate also increases with increasing SiO$_2$ thickness at the same bias current.
  Hence, it is a combination of bias current and material combination that leads to higher slew rates and, consequently, to an improved timing jitter.}
  \label{fig:6}
\end{figure}

To investigate the fundamental relation between pulse shape and timing jitter further, we look at the dependence of the timing jitter relative to the slew rate (ratio between pulse height and rise time) shown in Fig. \ref{fig:6}(a).

For an increased slew rate we observe a decreased timing jitter in agreement with D.Zhu\cite{Zhu19}.
It decreases rapidly from \SIrange{180}{100}{\pico\second} when the slew rate increases from \SIrange{250}{500}{\volt / \micro\second}.
Subsequently, the decrease slows down over \SI{50}{\pico\second} at \SI{1250}{\volt / \micro\second} until it reaches a timing jitter around \SI{25}{\pico\second} for a slew rate of \SI{3000}{\volt / \micro\second}.
Again, these findings are independent of the multi-layer structure used.
Importantly, only the detectors fabricated on thick SiO$_2$ layers, wafer and small gold mirrors allow to achieve high slew rates and therefore low timing jitter values.
The clear dependence of the timing jitter on the slew rate allows to deduce the timing jitter only by measuring the slew rate.
Thus, by measuring a single electrical pulse (e.g. pulse height and rise time) it is possible to predict the timing jitter within a slew rate dependent tolerance using CW excitation, eliminating the need of a pulsed laser for jitter measurements if a rough estimate of the timing jitter is sufficient. Hence, such a scheme can help to speed up the characterization process for many detectors after a given setup has been calibrated. Fig. \ref{fig:6}(b) shows a more detailed view of selected points at a bias current of \SI{12}{\micro A} and \SI{16}{\micro A}.
For an increased bias current, the slew rate becomes higher and the timing jitter lower.
Note that only the slew rate itself is strongly dependent on the multi-layer structure used.
We conclude that a combination of bias current and material combination leads to higher slew rates and thus improved timing jitter.
This fits well with the results in Fig. \ref{fig:3} and \ref{fig:4}(a), where a decreased rise time was always accompanied by an increased pulse height resulting in a higher slew rate.

\section{Conclusions}

In summary, we investigated the relation between the SNSPD timing jitter and the detector response at a wavelength of \SI{850}{\nano\meter} measured at \SI{4.5}{\kelvin} as well as the detection efficiency at \SI{780}{\nano\meter}.
The detectors were fabricated on a full-chip sized gold mirror (\SI{10}{\nano\meter} Ti / \SI{50}{\nano\meter} Au) with varying SiO$_2$ layer thicknesses, on small (\SI{100}{\micro\meter} diameter) gold mirrors with \SI{105}{\nano\meter} SiO$_2$, and on Si/SiO$_2$ wafers.
Besides a strong dependence on the bias current with an improved jitter for an increased current, we observed a dependence of the timing jitter on the multi-layer structure underneath the SNSPD.
In particular, the entire pulse shape changes depending on the multi-layer structure used at a fixed bias current of \SI{16}{\micro A}. This results in an increased pulse height with an increased SiO$_2$ thickness as well as for detectors on small gold mirrors or directly on a Si/SiO$_2$ wafer.
Additionally, we investigate the rise time and timing jitter for the different material combinations at the same bias current. We observed that with an increased SiO$_2$ thickness and for small gold mirrors the rise time decreases leading to an improved timing jitter.
Note that the origin of this is a capacitive behavior between the FCS gold mirror and the gold contact pads, which can be improved by using a thicker SiO$_2$ layer leading to a higher pulse height and faster rise time.
However, by fabricating the detector on a small gold mirror (\SI{100}{\micro\meter} diameter) underneath the detector (and not the entire Ti/Au contact pad) it is possible to restore the timing jitter as found on bulk Si/SiO2 substrates while maintaining the detection efficiency.
Hence, to combine a good efficiency with a good timing jitter either a small gold mirror or a thick SiO$_2$ layer ($> \SI{300}{\nano\meter}$) has to be used.
The second major finding is the relation between the timing jitter and the rise time.
In particular, we were able to extend these findings to the dependence of the timing jitter on the slew rate describing the ratio between the pulse height and rise time.
We found a general trend of the timing jitter to the slew rate, which is independent of the multi-layer structure used. We conclude that this enables us to estimate the timing jitter for a given detector pulse slew rate if a rough estimate of the timing jitter is sufficient. Thus, by only measuring single detector pulses this technique allows it to speed up the characterization process and paves the way for industrially scalable timing jitter measurements for instance with CW lasers.

\section*{Author Contributions}

R.F. designed study, analyzed data and drafted manuscript. R.F., L.Z., C.S. and S.S. fabricated samples. R.F. and C.S. performed the experiments. F.F. contributed simulations. J.F. and K.M. led the research projects. All authors discussed results and revised the manuscript.

\section*{Conflicts of interest}
There are no conflicts to declare.

\section*{Acknowledgements}
We gratefully acknowledge the German Federal Ministry of Education and Research via the funding program Photonics Research Germany (contract number 13N14846), via the funding program quantum technologies - from basic research to market (contract numbers 16K1SQ033, 13N15855, 13N15982, 13N16214 and 13N15760), and via the projects Q.com (contract number 16KIS0110) and MARQUAND (contract number BN105022), as well as the Deutsche Forschungsgemeinschaft (DFG, German Research Foundation) under Germany’s Excellence Strategy – EXC-2111 – 390814868.





\bibliography{references} 
\bibliographystyle{spiebib} 

\end{document}